\documentstyle[preprint,aps,prb]{revtex}
\begin{document}
\draft
\tightenlines 
\title{Far infrared giant dipole resonances in neutral quantum dots}
\author{Alain Delgado\cite{alain}$^1$, Lester Lavin\cite{lester}$^1$, 
 Roberto Capote\cite{capote}$^1$,  and Augusto Gonzalez
 \cite{augusto}$^{2,3}$}
\address{$^1$Centro de Estudios Aplicados al Desarrollo Nuclear,
 Calle 30 No 502, Miramar \\ La Habana, Cuba}
\address{$^2$Instituto de Cibernetica, Matematica y Fisica Calle E 
 309, Vedado, Habana 4, Cuba}
\address{$^3$Departamento de Fisica, Universidad de Antioquia, AA 1040, 
 Medellin, Colombia}
\date{Received: \today}

\maketitle

\begin{abstract}
A resonance behaviour of the far infrared absorption probability
at a frequency $\sim N^{1/4}$ is predicted for 
clusters of $N$ electron-hole pairs  ($2\le N\le 110$) 
confined in disk-shaped quantum dots. For radially
symmetric dots, the absorption is dominated by a Giant
Dipole Resonance, which accounts for more than 98 \%
of the energy-weighted photoabsorption sum rule.  
\end{abstract}
\vspace{.5cm}

\section{Introduction}

A well-established result of Nuclear Physics is the presence 
of giant resonance modes related to collective excitations in 
nuclei. Particularly, collective states corresponding to giant 
electric dipole oscillations 
are known in nuclei long ago \cite{gdr,gdrrev}. They show themselves 
as very high peaks in photoabsorption at energies $\sim~ 80
A^{-1/3}$ MeV, where A is the mass number. An understanding 
of this phenomenon can be achieved by means of a simple 
hydrodynamic model in which one considers small density 
perturbations against the background nuclear density and 
computes the small-oscillation frequencies associated to 
dipole excitations \cite{hydro}. A second, microscopic, approach
makes use of the nuclear shell model \cite{TD}. It is shown that
the giant dipole resonance (GDR) is a state coming from the 
splitting of a degenerate set of $1^-$ states, which takes 
almost the whole strength of the $1^-$ transition.

Collective states very similar to the GDR have been extensively
studied, both experimentally and theoretically, in Atomic
Physics. They are at the base of giant and broad atomic 
resonances in the photoionisation continuum \cite{Conn}. 
The photoabsorption cross section of metallic clusters
\cite{Conn,PWK91} exhibits also the dominance of a collective
oscillation mode at 2 - 3 eV \cite{BB94}. Theoretical calculations
based on the Random Phase Approximation (RPA) \cite{Conn,Lipp}
are shown to reproduce well the resonance position and dipole 
strength distribution.

In the present paper we show that, in electron-hole systems 
confined in a quantum dot, the energy-weighted photoabsorption
sum rule is almost saturated by GDR-like states, in which the 
electronic cloud oscillates relatively to the hole cloud. This result
is mainly related to the existence of positively and negatively
charged particles, and not to a particular dot geometry. Thus, 
we will use a very simplified model of disk-shaped dot, in which
the lateral confinement is ideally parabolic \cite{WHFJ96}.

$N$ electron-hole pairs are supposed to be created in the dot 
by means of, e. g., laser pumping with a typical $\sim$ 2 eV energy,
as corresponding to a semiconductor band gap. The created particles live 
hundreds of picoseconds and even more \cite{BG93}. We will study
the linear absorption of a second far infrared (FIR), $\sim$ 5 meV, 
light wave by the $N$-pair cluster. The period of this signal is 
$\sim 10^{- 13}$s, thus the cluster may be considered as stable when 
studying absorption. We will use the dipole 
approximation for the interaction hamiltonian. 
 
Experiments on FIR study of excitonic systems have been conducted for 
a long time \cite{FIR}. More recently, the modifications of the 
interband photoluminescence (PL) of quantum well excitons caused by 
the absorption of FIR radiation have been studied \cite{ODR} by means of 
a technique baptised as ODR (optical detection of FIR resonances). These
experiments are conducted at relatively small exciton densities, where 
many-body effects are not expected to play a significant role. To study
the collective FIR absorption by a multi-exciton system, intense laser
beams for the interband processes shall be used. Experimentally, 
densities as high as 10$^{12}$ electron-hole (e-h) pairs/cm$^2$ in GaAs
wells have been achieved by using pulsed (5 ps long) dye lasers 
\cite{Wolfe}. The effective absorbed power is greater than 2 MW/cm$^2$.

To our knowledge, however, the effect of FIR radiation on dense excitonic
systems have not been experimentally studied so far, although intense
enough laser beams and the ODR technique are available. For simplicity 
of the calculations, we confine the system in a quantum dot (qdot). This 
confinement makes the detection a little more cumbersome, but still
experimentally achievable. Single qdot multi-excitonic PL has been
isolated from the averaged PL by means of microscopy techniques
(see, for example, Ref. [\onlinecite{experiments}]). Thus, in principle, 
the ODR study of single qdots is also possible.

The organisation of the paper is as follows. In Section II, we present the
model to be used in the paper. Next, the basic RPA equations are explicitly
written in Section III. In the later two sections, the main results and a 
few concluding remarks are given.

\section{The model}

We will study a quasi-planar quantum dot, i. e. very narrow in the growth
direction (the $z$ axis), in such a way that the first quantum well 
sub-band approximation works, and no significant modifications of 
Coulomb interactions due to $z$-averaging arise. A direct band-gap material
and high barriers for both electrons and holes in the $z$ direction are
assumed to exist. Confinement makes also the hole sub-bands to split. Thus,
only one e and one h bands will be considered. e-h exchange will be 
neglected. As mentioned above, the lateral confinement will be modelled
by an ideal parabolic potential. Finally, for simplicity, we take 
$m_e=m_h$. This is a non realistic assumption, justified only by the 
absence of experimental results. In a GaAs dot, for example, 8 nm wide
in the $z$ direction, the ratio of the in plane heavy hole to electron 
masses is around 2. The equal-mass simplification makes calculations 
more easy, giving nevertheless a qualitatively correct picture of what is 
happening. 

The second-quantised Hamiltonian for the two-dimensional motion of 
electrons and holes is written in oscillator units as

\begin{eqnarray}
H&=&\sum_i \epsilon_i (\hat e_i^{\dagger} \hat e_i+
  \hat h_i^{\dagger} \hat h_i)
 +\frac{\beta}{2} \sum_{i,j,k,l}\langle i,j|1/r|k,l\rangle
   \hat e_i^{\dagger}\hat e_j^{\dagger}\hat e_l\hat e_k\nonumber\\
 &&+\frac{\beta}{2} \sum_{i,j,k,l}\langle i,j|1/r|k,l\rangle
   \hat h_i^{\dagger}\hat h_j^{\dagger}\hat h_l\hat h_k
   -\beta \sum_{i,j,k,l}\langle i,j|1/r|k,l\rangle
   \hat h_i^{\dagger}\hat e_j^{\dagger}\hat e_l\hat h_k,
 \end{eqnarray}

\noindent
where $\hat e_i$($\hat h_i$) are electron (hole) operators, 
$\beta=\sqrt{(\frac{m_e e^4}{\kappa^2\hbar^2})/(\hbar\omega)}$,
$\kappa$ is the dielectric constant, and $\omega$ is the dot 
frequency. The Coulomb matrix elements are defined as

\begin{eqnarray}
\langle i,j|1/r|k,l\rangle &=&\delta (s_{zi},s_{zk}) 
 \delta(s_{zj},s_{zl})\int \frac{{\rm d^2}r_1 {\rm d^2}r_2}
 {|\vec r_1-\vec r_2|}
 \phi_i^*(\vec r_1) \phi_j^*(\vec r_2)\phi_k (\vec r_1) \phi_l 
 (\vec r_2),
\end{eqnarray}

\noindent
and the explicit form of the harmonic-oscillator orbitals is 
the following

\begin{equation}
\phi_j=C_{n_j,m_j} r^{|m_j|} L_{n_j}^{|m_j|}(r^2)e^{-r^2/2}
 e^{i m_j \theta},
\end{equation}

\noindent
where $n_j$ and $m_j$ are oscillator quantum numbers, and 
$C_{n,|m|}=\sqrt{n!/[\pi~(n+|m|)!]}$. The one-particle energies 
in the quadratic potential are $\epsilon_j=1+2 n_j+|m_j|$. 

We consider the dot in the strong-confinement regime, where the 
one-particle energy dominates the total energy. It means, $\beta<\beta_c$,
where $\beta_c$ may be estimated \cite{GQRCR98} from the breakdown of 
perturbation theory or from the onset of BCS pairing to be 
$\beta_c\approx 0.58~N^{1/4}$. The associated value of
density is $\rho_c\approx 4/(\pi a_B^2)$. For GaAs, the Bohr radius is
$a_B\approx 14$ nm, thus $\rho_c\approx 5\cdot 10^{11}$ pairs/cm$^2$.

For simplicity, we study closed-shell quantum dots. The number
of electrons is equal to the number of holes, and is given 
by the expression $N=N_s (N_s+1)$, where $N_s$
is the last harmonic-oscillator filled shell. The zeroth-order 
ground state-function is a
product of Slater determinants for electrons and holes. The total 
angular momentum, total electron spin and total hole spin
are all equal to zero for this state.

The lowest excitations against the ground state are one-particle
excitations from the last filled shell to the next empty shell. 
The states with total angular momentum, $M=\pm 1$, and electron 
and hole total spins, $S_e=S_h=0$, are the candidates among which 
we shall look for the GDR. To be definite, we will consider $M=1$.
Notice that there are $N_s$ 
orbitals in the last shell, that is $2 N_s$ electrons and,
thus, $2 N_s$ wave functions with $M=1$. There are also $2 N_s$
functions corresponding to hole excitations. It makes a total of  
$4 N_s$ many-particle states.

The Coulomb interaction breaks the oscillator degeneracy. The 
collective nature of the GDR is manifested in two facts: 
({\it i}) it acquires the highest excitation energy in the 
multiplet, and ({\it ii}) the electric dipole transition probability 
from the ground state is strongly enhanced \cite{TD}. We 
will present RPA calculations in support of these statements.

\section{The random phase approximation}

In the RPA \cite{RR80}, we allow a general correlated 
ground state, $|RPA\rangle$, and the excited states are looked
for in the form

\begin{equation}
\Psi = Q^{\dagger} |RPA\rangle ,
\end{equation}

\noindent
where the $Q^{\dagger}$ operator is given by the expression

\begin{equation}
Q^{\dagger}=\sum_{\sigma,\lambda}(X^e_{\sigma\lambda} 
 e^{\dagger}_{\sigma} e_{\lambda}+X^h_{\sigma\lambda}
 h^{\dagger}_{\sigma} h_{\lambda}-Y^e_{\sigma\lambda} 
 e^{\dagger}_{\lambda} e_{\sigma}-Y^h_{\sigma\lambda} 
 h^{\dagger}_{\lambda} h_{\sigma}).
\end{equation}

\noindent
The index $\lambda$ runs over the last Hartree-Fock (HF) filled shell, 
and $\sigma$ runs over the first unfilled shell. The $X$ coefficients
are nonzero only for ``up'' transitions, i.e. for which the angular 
momentum values satisfy $m_{\sigma}-m_{\lambda}=1$, and the
$Y$ coefficients are nonzero only for ``down'' transitions, i.e. $\lambda$ 
and $\sigma$ satisfying $m_{\lambda}-m_{\sigma}=1$. They are obtained from 
the RPA equations

\begin{eqnarray}
\left(\matrix{A &B\cr
-B^t &-A^d} \right) 
\left( \matrix{X\cr Y} \right)=\Delta E 
\left( \matrix{X\cr Y} \right),
\label{RPAeq}
\end{eqnarray}

\noindent
in which the $A$ matrix is given by

\begin{eqnarray}
\left(\matrix{A^{ee} &A^{eh}\cr
A^{he} &A^{hh}} \right),
\end{eqnarray}
 
\noindent
and its matrix elements are the following

\begin{eqnarray}
A^{ee}_{\sigma\lambda,\tau\mu} &=& (\epsilon^{HF}_{\sigma}-
 \epsilon^{HF}_{\lambda}) \delta_{\sigma\tau} \delta_{\lambda\mu}
 + \beta \left( \langle \sigma,\mu |1/r|\lambda,\tau\rangle -\langle 
 \sigma,\mu |1/r|\tau,\lambda\rangle \right),\label{matrixA1}\\
A^{eh}_{\sigma\lambda,\tau\mu} &=& -\beta\langle \sigma,\mu |
 1/r|\lambda,\tau \rangle .
\label{matrixA2} 
\end{eqnarray}

\noindent
$A^{hh}=A^{ee}$, $A^{he}=A^{eh}$ because of the electron-hole 
symmetry of our hamiltonian. $\tau$ and $\mu$ are indexes 
similar to $\sigma$ and $\lambda$, respectively.
The $B$ matrix has a similar structure. Its matrix elements, calculated
between up and down transitions, are given by

\begin{eqnarray}
B^{ee}_{\sigma\lambda,\tau\mu} &=& \beta ( \langle \sigma,\tau
|1/r|\lambda,\mu\rangle -\langle \sigma,\tau |1/r|
\mu,\lambda\rangle ),\\
B^{eh}_{\sigma\lambda,\tau\mu} &=& -\beta\langle \sigma,\tau 
|1/r|\lambda,\mu \rangle. 
\end{eqnarray}

\noindent
$B^{hh}=B^{ee}$ and $B^{he}=B^{eh}$. $B^t$ is the transpose of
the numerical matrix $B$, and the $A^d$ matrix is 
calculated from (\ref{matrixA1},\ref{matrixA2}) between down transitions.
Usually, positive (physical) and
negative (unphysical) excitation energies of equal magnitude
come from (\ref{RPAeq}). The physical solutions annihilate the
RPA ground state

\begin{equation}
Q |RPA\rangle = 0,
\end{equation}

\noindent
and satisfy the normalisation condition

\begin{equation}
1= \sum_{\sigma,\lambda} 
 \{ (X_{\sigma\lambda})^2-(Y_{\sigma\lambda})^2\} .
\end{equation}
 
The FIR radiation to be absorbed is assumed to be linearly polarised,
with the electric field vibrating along the $y$ direction.
The dipole matrix elements are computed from

\begin{equation}
\langle \Psi |D| RPA\rangle = \sum_{\sigma,\lambda} 
 \{ (X^h_{\sigma\lambda}-X^e_{\sigma\lambda}) \langle\sigma|y|
 \lambda\rangle +
    (Y^h_{\sigma\lambda}-Y^e_{\sigma\lambda}) \langle\lambda|y|
 \sigma\rangle \} .
\end{equation}

The calculated dipole elements fulfil the energy-weighted
sum rule \cite{RR80}

\begin{eqnarray}
\sum_{\Psi}\Delta E~|\langle \Psi |D|RPA \rangle|^2=
 \sum_{\Psi_0}\Delta E_0 |\langle \Psi_0 |D|0 \rangle|^2=N/2,
\label{S1}
\end{eqnarray}

\noindent
where 0 indexes refer to the noninteracting $\beta=0$ case.

The HF energies and one-particle wave functions were 
obtained selfconsistently by means of an iterative scheme. 
We write

\begin{equation}
|\alpha \rangle = \sum_i C_{\alpha i} |i\rangle ,
\end{equation}

\noindent
where latin indexes refer to harmonic oscillator orbitals. 
272 orbitals, i. e. 16 harmonic oscillator shells where included 
in the calculations. The HF equations read

\begin{equation}
\sum_j \{\epsilon_i \delta_{ij}-\beta \sum_{\mu\le F}
 \sum_{k,l} C_{\mu k} C_{\mu l} \langle i,k|1/r|l,j\rangle \}
 C_{\alpha j} = \epsilon_{\alpha} C_{\alpha i} ,
\end{equation}

\noindent
and the starting approximation is $C_{\alpha i}=\delta_
{\alpha i}$. $F$ denotes the Fermi level. A few iterations are enough to 
reach convergence at very low $\beta$ values, while at $\beta\sim 1$, 
15 iterations are needed.

\section{Results}

Excitation energies and dipole strength distributions were
computed for $N$ ranging between 2 and 110, and $\beta$ in
the interval $0\le\beta\le 0.6~N^{1/4}$. For larger values
of $\beta$, pairing becomes very important \cite{GQRCR98},
and should be taken into account more properly. Let us
stress, however, that pairing should reinforce the picture
of GDR dominance of photoabsorption, leading to an
increase of the excitation energy and, consequently, to a
decrease in the same proportion of the dipole element squared.

The results for $N=110$ pairs at $\beta=1$ are shown 
in Fig. \ref{fig1}, where dipole matrix elements squared vs
excitation energies are shown. The fraction of the 
sum rule (\ref{S1}) accounted for the GDR state
is close to 100 \%. Notice that only collective states,
for which $X^e=-X^h$, are represented in the 
figure. The rest of the states show a threefold degeneracy 
and give practically zero dipole matrix elements. Let us note that
the energy gap is increased ($\Delta E>1$) after the Coulomb interaction
is switched on. This is a signal that the overall interaction among
particles is attractive.

We draw in Fig. \ref{fig2}a the GDR excitation energies at $\beta=1$
as a function of $N^{1/4}$. Notice the approximate dependence
$\Delta E_{gdr}-1\approx 0.7 +0.4~N^{1/4}$. At $\beta\ne 1$, the dependence
is $\Delta E_{gdr}-1\approx (0.7 +0.4~N^{1/4})\beta$. 
Turning back to ordinary units, we get an excitation energy

\begin{equation}
\Delta E_{gdr}\approx (1+0.7 \beta)\hbar\omega+0.4\frac{\hbar^2}
 {m a_B} \sqrt{2\pi/3}~\rho^{1/2},
\end{equation}

\noindent
where $\rho\approx 3 N^{1/2}m\omega/(2\pi\hbar)$ is 
the density of pairs. For
large enough $N$, and $\omega\to 0$ in order to preserve a 
constant density, $\Delta E_{gdr}\approx 0.6~\hbar^2\rho^{1/2}/
(m a_B)$, leading for example to $\Delta E_{gdr}
\approx 4.5$ meV for typical values of the constants in GaAs 
and $\rho\approx 5\cdot 10^{11}$ pairs/cm$^2$.

The GDR dipole elements squared at $\beta=1$ as a function of $N$ are
shown in Fig. \ref{fig2}b. The points fit a linear dependence 
$\sim 0.17~N$. For $\beta\ne 1$, the saturation of the sum rule implies
that $D^2\approx N/(2\Delta E_{gdr})\approx N/(2+(1.4+0.8 N^{1/4})\beta)$.

Finally, we draw in Fig. \ref{fig3} the fraction of the energy-weighted
sum rule (\ref{S1}), the $S1$ sum rule, corresponding to the transition
to the GDR state. This fraction is higher than 98\% for any of the 
systems shown in Figs. \ref{fig3}a and \ref{fig3}b. (We got a slight
violation, fraction=1.01, for $N=110$ at $\beta=1$, but this is an
artifact related
to the finiteness of the basis used to construct the HF states).   

\section{Conclusions}

We have shown that the far infrared absorption spectrum of neutral 
confined systems of electrons and holes is dominated by a giant dipole 
resonance which accounts for more than 98\% of the energy-weighted
photoabsorption sum rule. Calculations were done in a very simplified
two-band model with $m_e=m_h$ and a disk-shaped parabolic dot. The 
qualitative conclusions are, however, expected to be valid for realistic
systems since they are mainly related to the existence of positively and
negatively charged particles in the system, which causes the
enhancement of dipole oscillations.  

For the resonance position at intermediate $N$, we obtained the fit, 
$\Delta E_{gdr}=\{1+(0.7+0.4~N^{1/4})\beta\}\hbar\omega$. In terms
of the density, we obtain $\Delta E_{gdr}\approx (1+0.7 \beta)\hbar\omega+
0.4 \sqrt{2\pi/3}~\hbar^2\rho^{1/2}/(m~a_B)$. The dipole strength scales 
with $N$: $D^2\sim N$, at small $N$ values, and $D^2\sim N^{3/4}$ at 
larger values of $N$. 

The GDR state we found is the analogue in a finite two-dimensional 
system of the out-of-phase plasma frequency in the two-component plasma
\cite{Pines}. The term proportional to $\hbar\omega$ in the excitation
energy is a shell effect of the external harmonic-oscillator potential.

We worked in the strong-confinement regime, $\beta < 0.6~N^{1/4}$. For
still higher values of $\beta$, BCS pairing should be explicitly included
in the calculations. Recently, we have computed the ground-state energy
of the confined $N$-pair system by using a BCS trial function
\cite{GQRCR98}. The effect of pairing in $\Delta E_{gdr}$ could be taken
into account by means of a quasiparticle RPA scheme \cite{RR80}, which 
may be taken as the analogue for a finite system of the Gor'kov's pairing
approximation in the excitonic insulator phase \cite{KM65,JRK67,CG88}. At
relatively high densities
we expect, qualitatively, a shift of the resonance position in
accordance to the opening of the gap at the Fermi level:

\begin{equation}
\Delta E_{gdr}=\sqrt{(2 \Delta_F)^2+\left((1+0.7 \beta)\hbar\omega+
0.4 \sqrt{2\pi/3}~\hbar^2\rho^{1/2}/(m~a_B)\right)^2}, 
\end{equation}

\noindent
where $\Delta_F$ is the gap function at the Fermi energy. The dipole
strength shall decrease in the same proportion, as dictated by (\ref{S1}).

\acknowledgements 
A. D. and R. C. acknowledge support from the Universidad de 
Antioquia (where part of this work was done) and from 
the Caribbean Network for Theoretical Physics.

\begin{figure}
\caption{Dipole matrix elements squared for 110 pairs at 
 $\beta=1$.}
\label{fig1}
\end{figure}

\begin{figure}
\caption{a) GDR excitation energy as a function of the 
 number of pairs in the dot at $\beta=1$.\\
 (b) GDR dipole matrix elements squared as a 
 function of $N$ at $\beta=1$.}
\label{fig2}
\end{figure}

\begin{figure}
\caption{Fraction of the energy-weighted sum rule accounted for the
GDR: a) As a function of $\beta$ for $N=110$. b) As a function of 
the number of pairs at $\beta=1$.}
\label{fig3}
\end{figure}

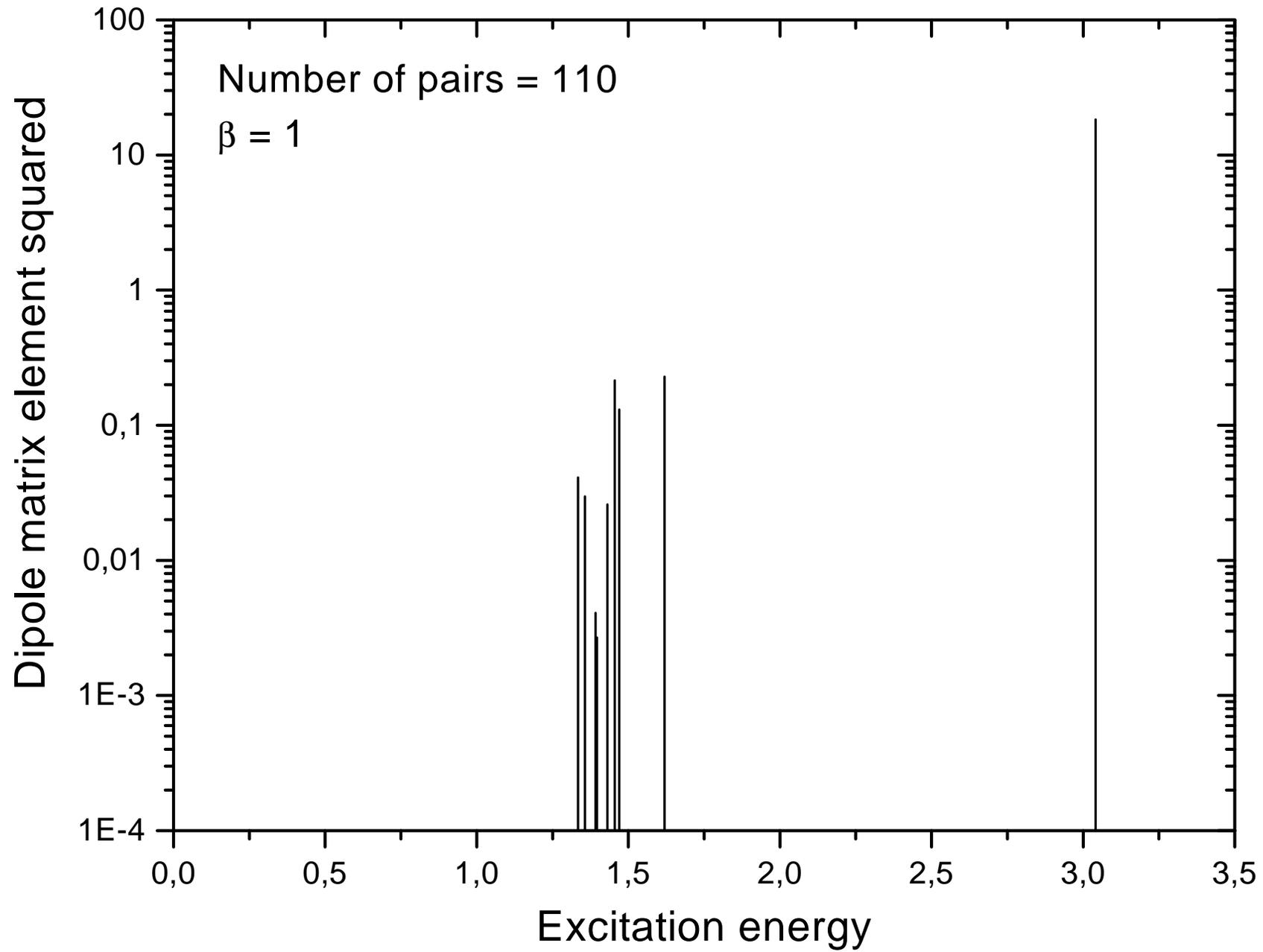
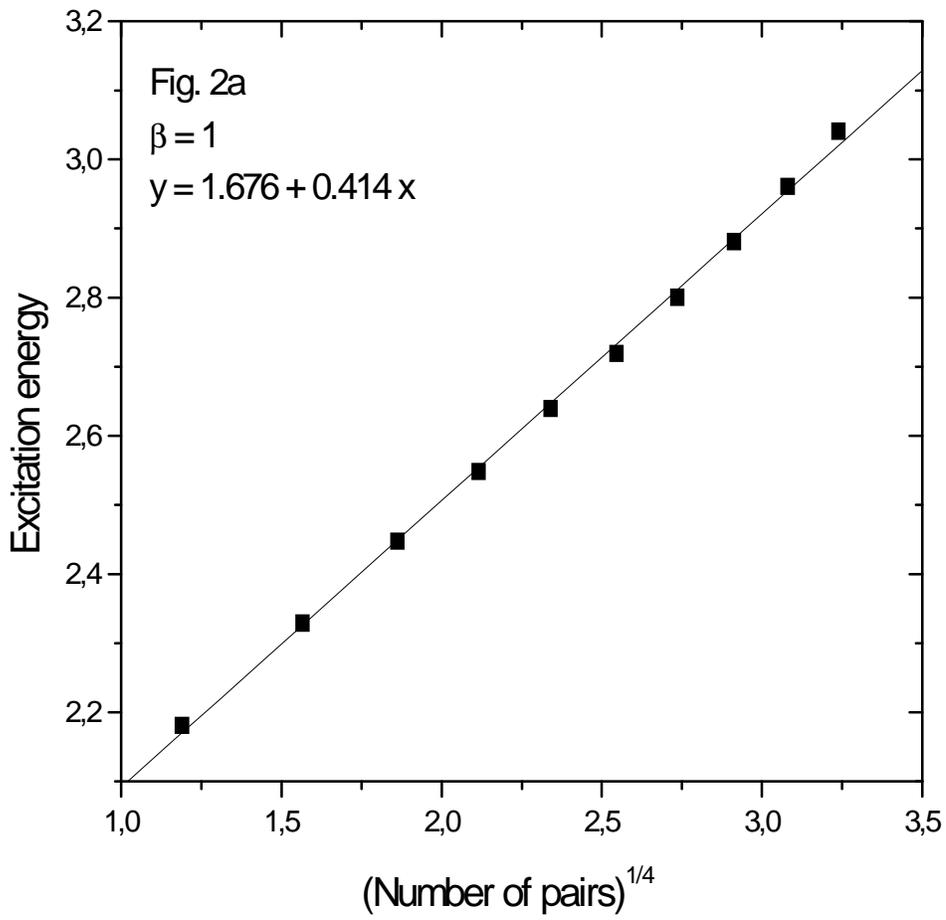
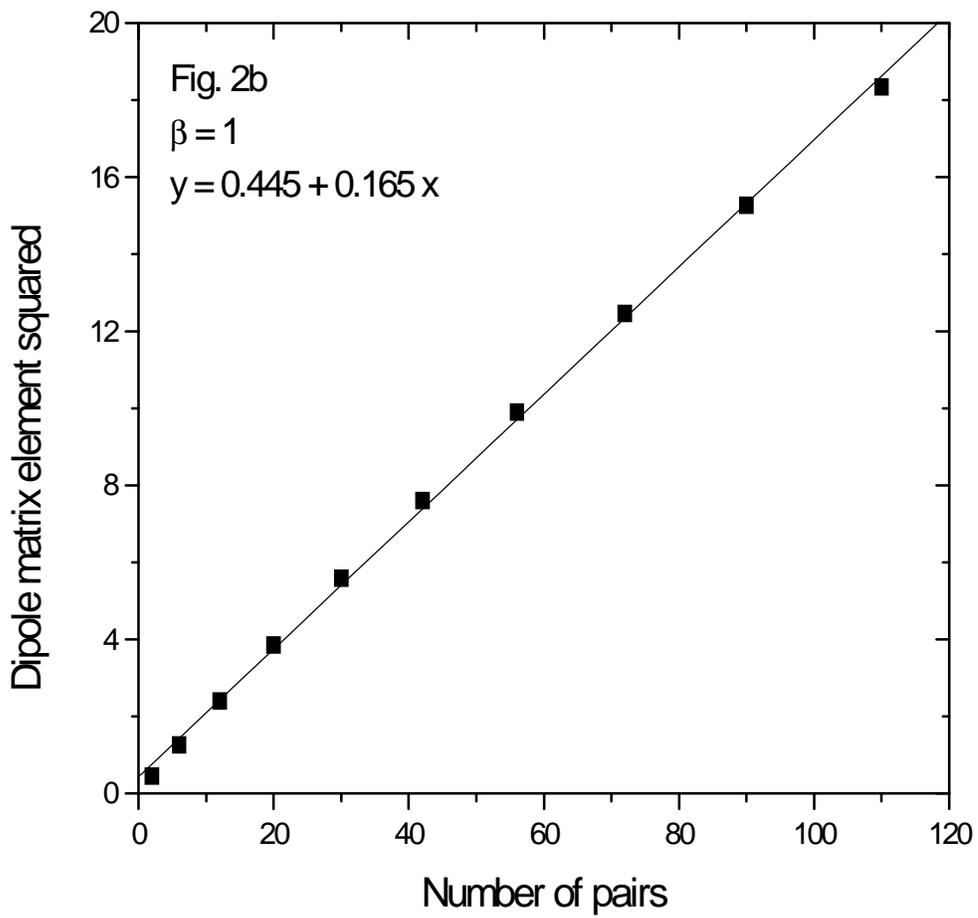
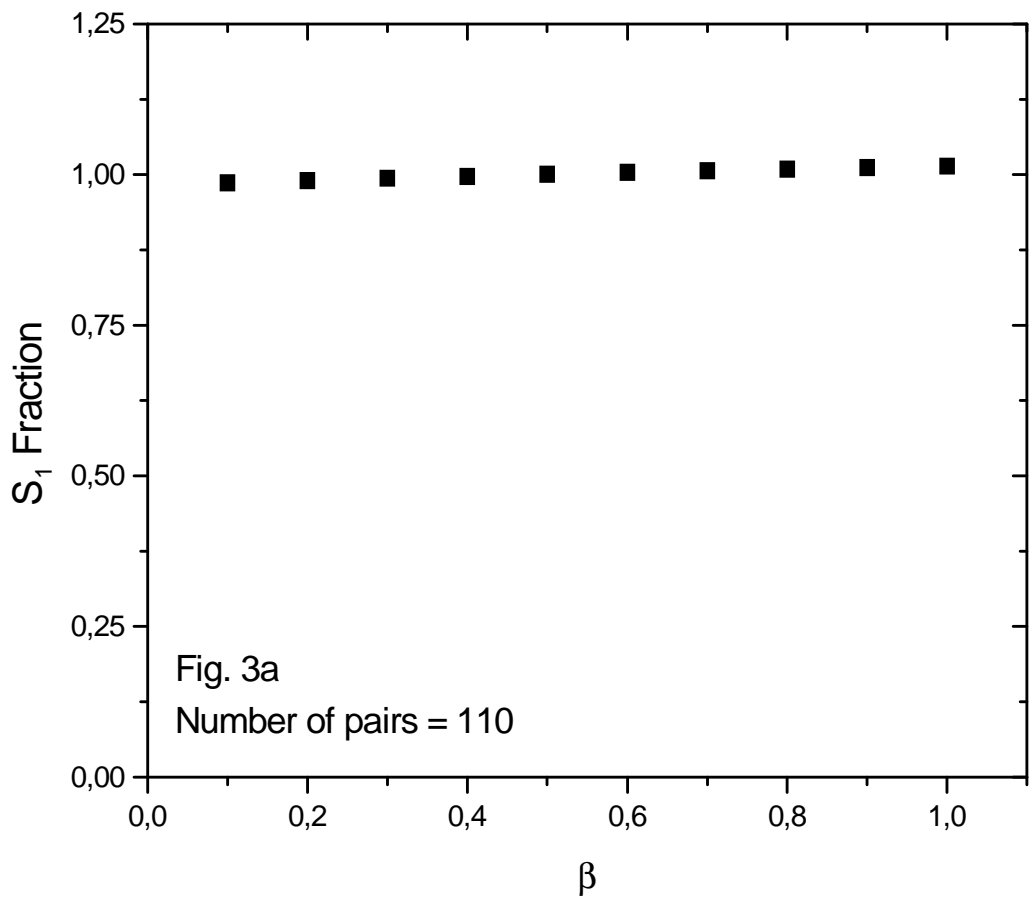
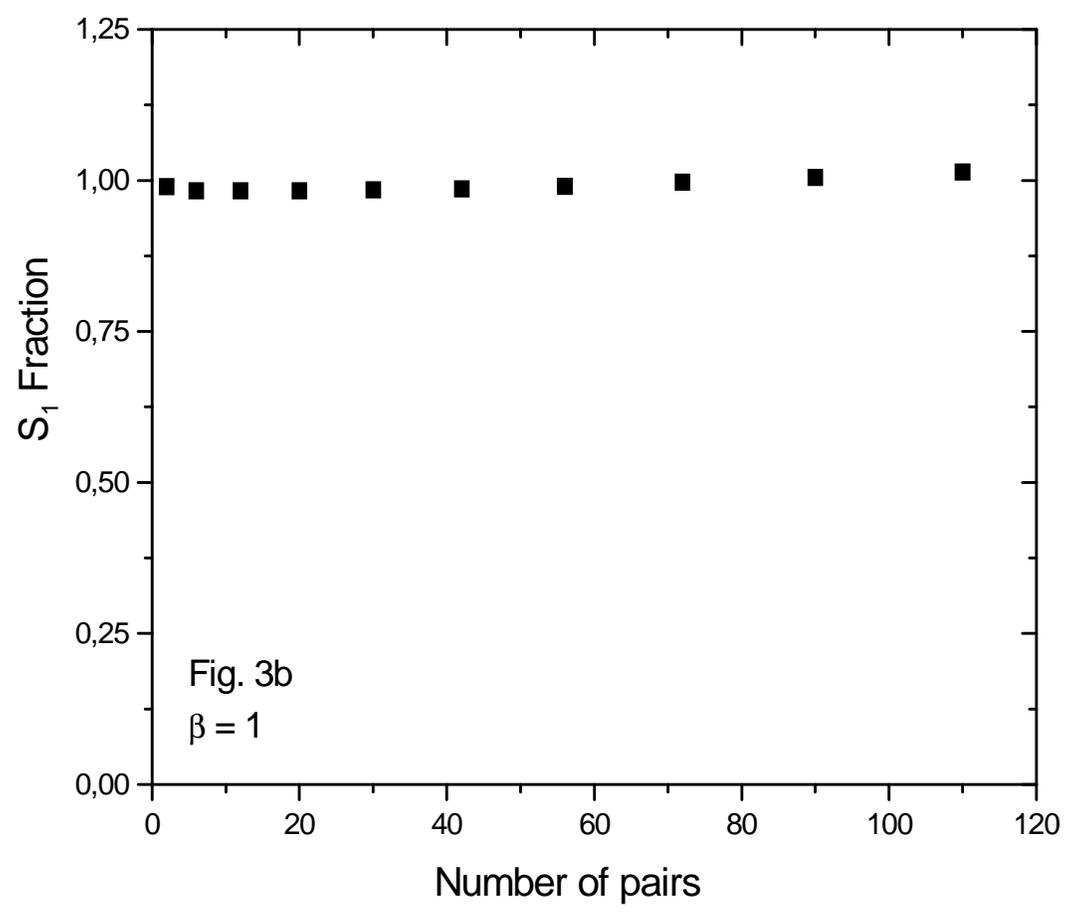
\end{document}